\documentstyle[multicol,aps]{revtex}
\def\beq{\begin{eqnarray*}}
\def\eeq{\end{eqnarray*}}
\def\beqn{\begin{eqnarray}}
\def\eeqn{\end{eqnarray}}

\begin{document}
\title{Vacuum fluctuations, accelerated motion\\
and conformal frames}
\author{Marc-Thierry JAEKEL$^{(a)}$ and Serge REYNAUD$^{(b)}$}
\address{(a) Laboratoire de Physique Th\'{e}orique de l'ENS\thanks{%
Laboratoire du CNRS associ\'{e} \`{a} l'Ecole Normale Su\-p\'{e}\-rieu\-re et
\`{a} l'Uni\-ver\-si\-t\'{e} Paris-Sud.},\\
24 rue Lhomond, F75231 Paris Cedex 05, France\\
(b) Laboratoire Kastler Brossel\thanks{%
Laboratoire de  l'Ecole Normale Su\-p\'{e}\-rieu\-re et de
l'Uni\-ver\-si\-t\'{e} Pierre
et Marie Curie associ\'{e} au CNRS.}, case 74,\\
4 place Jussieu, F75252 Paris Cedex 05, France}
\date{{\sc Quantum and Semiclassical Optics} {\bf 7} 499-508 (1995)}
\maketitle

\begin{abstract}
Radiation from a mirror moving in vacuum electromagnetic fields is shown to
vanish in the case of a uniformly accelerated motion. Such motions are
related to conformal coordinate transformations, which preserve correlation
functions characteristic of vacuum fluctuations. As a result, vacuum
fluctuations remain invariant under reflection upon a uniformly accelerated
mirror, which therefore does not radiate and experiences no radiation
reaction force. Mechanical effects of vacuum fluctuations thus exhibit an
invariance with respect to uniformly accelerated motions.

\end{abstract}

\begin{multicols}{2}

\section{Introduction}

Quantum fields carry energy and momentum and induce mechanical effects upon
mirrors. As known for a long time \cite{Einstein}, a mirror immersed in
thermal fluctuations experiences a mean dissipative force which tends to
damp it to a null velocity and which is connected through
fluctuation-dissipation relations to force fluctuations experienced by a
motionless mirror. As a result of these forces, quantum field fluctuations
induce a Brownian motion for the mirror's position. Such effects persist for
mirrors immersed in vacuum fluctuations, i.e. at the limit of zero
temperature \cite{QO92,Barton,JP93a}.

In particular, a mirror with an arbitrary motion in vacuum experiences a
radiation reaction force, which can be analyzed within the framework of
Quantum Field Theory \cite{FullingD,FordV}. Being treated as a moving
classical boundary for the fields, the mirror modifies the scattered fields
and leads to a radiation of energy and momentum which induces a radiation
reaction force. This force may also be analyzed in the mirror's proper frame
and, then, seen to result from the modification of vacuum fluctuations under
a frame transformation fitting the mirror's motion \cite{QO92}. The force
vanishes for a motion with a uniform velocity, since vacuum fluctuations are
invariant under Lorentz coordinate transformations and, thus, preserved by
reflection upon a mirror with a uniform velocity. In other words, the
radiation reaction force possesses the symmetry properties required by
relativity of motion in vacuum.

The case of uniformly accelerated motion requires a careful examination, in
connection with the question of whether or not vacuum fluctuations appear
different after coordinate transformations to accelerated frames \cite{Unruh}%
{}.

For a perfect mirror moving in vacuum scalar fields in a two-dimensional
space-time \cite{FullingD}, the radiation reaction force is proportional to
the Abraham-Lorentz derivative and, therefore, vanishes for uniformly
accelerated motion \cite{Born}. The latter property still holds for plane
mirrors moving in a four-dimensional space-time, in conformally invariant
scalar fields \cite{FordV} or electromagnetic fields \cite{Maia} (in these
two cases, the forces have been evaluated only for small displacements of
the mirrors).

In the present paper, we show quite generally that the absence of radiation
from uniformly accelerated mirror, and therefore the absence of radiation
reaction force, result from the invariance of vacuum fluctuations of such
quantum field theories under conformal coordinate transformations.

We shall thus define accelerated frames through those transformations which
are generated by Poincar\'e transformations and four-dimensional Minkowskian
inversion \cite{CunninghamB}, and which are known to fit \cite{Hill,Page}
the usual relativistic definition of uniformly accelerated motion \cite{Born}%
. Propagation equations for the electromagnetic field, i.e. Maxwell
equations, are preserved in such transformations \cite{CunninghamB}. This
property, now considered a particular case of conformal invariance of
electromagnetic field theory \cite{FultonRW}, implies that a conformal
factor, i.e. a scale factor for length or time measurements, cannot be
detected by electromagnetic probes \cite{MasshoonG}. Due to the close
connection between propagators and commutators in quantum theory, it may be
expected that field correlation functions characteristic of vacuum
fluctuations are also preserved by conformal transformations to accelerated
frames. We will confirm that this is indeed the case and, then, deduce the
absence of radiation by uniformly accelerated mirrors from this invariance
property.

After having established this result, we shall discuss its relation to the
commonly referred equivalence between accelerated vacuum and thermal state
\cite{Unruh}. This property is associated with a different choice for
coordinate transformations from inertial to accelerated frames, namely the
Rindler transformations defined as hyperbolic transformations preserving
rigidity of spatially extended bodies \cite{Rindler}. We will emphasize that
the result obtained here, i.e. the absence of radiation from uniformly
accelerated mirrors, only depends upon this motion, but not upon a
particular choice of coordinate transformations fitting it.

\section{Uniformly accelerated motion}

Uniformly accelerated motion of a point-like body is defined according to
the usual hyperbolic relation \cite{Born} between velocity $v^\mu $ and
proper time $\tau $, where the constant $a$ measures acceleration
\begin{eqnarray*}
v^\mu \left( \tau \right) =v^\mu \left( 0\right) {\rm ch}\left( a\tau
\right) +\stackrel{\cdot }{v}^\mu \left( 0\right) {\rm sh}\left( a\tau
\right)
\end{eqnarray*}
The velocity $v^\mu $ is defined with respect to proper time
\begin{eqnarray*}
v^\mu \left( \tau \right)  &=&\frac{dx^\mu }{d\tau } \\
d\tau ^2 &=&dx^2
\end{eqnarray*}
as well as its derivatives $\stackrel{\cdot }{v}^\mu $ and $\stackrel{\cdot
\cdot }{v}^\mu $
\begin{eqnarray*}
\stackrel{\cdot }{v}^\mu \left( \tau \right)  &=&\frac{dv^\mu }{d\tau } \\
\stackrel{\cdot \cdot }{v}^\mu \left( \tau \right)  &=&\frac{d^2v^\mu }{%
d\tau ^2}
\end{eqnarray*}
 From the property
\begin{eqnarray*}
v^2\left( \tau \right) =1
\end{eqnarray*}
one deduces that
\begin{eqnarray*}
v\left( \tau \right) .\stackrel{\cdot }{v}\left( \tau \right)  &=&0 \\
\stackrel{\cdot }{v}\left( \tau \right) ^2 &=&a^2
\end{eqnarray*}
Throughout the paper, scalar products, squares and index manipulation are
defined with respect to Minkowski metric $\eta _{\mu \nu }={\rm diag}%
(1,-1,-1,-1)$%
\begin{eqnarray*}
x.y &\equiv &\eta _{\mu \nu }x^\mu y^\nu =x_\nu y^\nu  \\
x^2 &\equiv &x.x
\end{eqnarray*}
The foregoing relations hold in particular for $\tau =0$ and thus constrain
the initial values $v^\mu (0)$ and $\stackrel{\cdot }{v}^\mu (0)$.

It follows that uniformly accelerated motions may be characterized by the
relation
\begin{eqnarray}
w^\mu \equiv \stackrel{\cdot \cdot }{v}^\mu +v^\mu \stackrel{\cdot }{v}^2=0
\label{eq1}
\end{eqnarray}
It may be noticed that the Abraham-Lorentz force, i.e. the radiation
reaction force experienced by a charge moving in electromagnetic vacuum, is
proportional to the quantity $w^\mu $ and thus vanishes for uniformly
accelerated motions \cite{Born}. This is also the case for the radiation
reaction force experienced by a perfectly reflecting point-like mirror
moving in vacuum scalar fields in a two-dimensional space-time \cite
{FullingD}.

An important property of equation (\ref{eq1}) is its invariance under
conformal coordinate transformations. Already implicitly remarked by Bateman
\cite{Bateman}, this was explicitly demonstrated by Hill \cite{Hill}. This
property, a key point in our approach, is summarized in the next section.

\section{Conformal accelerated frames}

Let us first recall the definition of conformal changes of frame. A
conformal metric tensor is proportional to Minkowski metric
\begin{eqnarray}
g_{\mu \nu }=\lambda \left( x\right) ^2\eta _{\mu \nu }  \label{eq2a}
\end{eqnarray}
and its Ricci curvature tensor $R_{\mu \nu }$ is given by
\begin{eqnarray*}
R_{\mu \nu } &=&-\eta _{\mu \nu }\eta ^{\alpha \beta }\left( \varphi
_{\alpha \beta }+2\varphi _\alpha \varphi _\beta \right) -2\left( \varphi
_{\mu \nu }-\varphi _\mu \varphi _\nu \right)  \\
\varphi _\mu  &=&\partial _\mu \left( \ln \left( \lambda \right) \right)  \\
\varphi _{\mu \nu } &=&\partial _\mu \varphi _\nu =\partial _\nu \varphi
_\mu
\end{eqnarray*}
For a conformal metric obtained from a Minkowski metric through a coordinate
transformation, curvature tensors vanish ($R_{\mu \nu }=0$), so that the
conformal factor $\lambda $ takes the form
\begin{eqnarray}
\lambda \left( x\right) =\frac \beta {1-2\alpha .x+a^2x^2}  \label{eq2b}
\end{eqnarray}
where $\alpha ^\mu $ and $\beta $ are five integration constants.

The coordinate transformations corresponding to a change from Minkowski
frame to the conformal frame (\ref{eq2a}) are thus identified as
transformations generated by Poincar\'{e} transformations and inversions
\cite{CunninghamB}. It is possible to come back from the conformal frame $%
{\cal R}$ (coordinates $x^\mu $ and metric tensor $g_{\mu \nu }$) to a
Minkowski frame $\overline{{\cal R}}$ (coordinates $\overline{x^\mu }$) by
combining an inversion, a translation and an inversion
\begin{eqnarray}
-\beta \frac{\overline{x}^\mu }{\overline{x}^2}=\alpha ^\mu -\frac{x^\mu }{%
x^2}  \label{eq3a}
\end{eqnarray}
that is
\begin{eqnarray}
\overline{x}^\mu =\lambda \left( x\right) \left( x^\mu -x^2\alpha ^\mu
\right)   \label{eq3b}
\end{eqnarray}
with $\lambda $ the conformal factor given in equation (\ref{eq2b}). The
metric tensor $\overline{g}_{\mu \nu }$ in the frame $\overline{{\cal R}}$
is Minkowskian
\begin{eqnarray*}
\overline{g}_{\mu \nu }=\frac{\partial x^\rho }{\partial \overline{x}^\mu }%
\lambda \left( x\right) ^2\eta _{\rho \sigma }\frac{\partial x^\sigma }{%
\partial \overline{x}^\nu }=\eta _{\mu \nu }
\end{eqnarray*}
To establish a property in the conformal frame ${\cal R}$, it is then
sufficient to have it verified in Minkowski frames and preserved by
translations and inversions
\begin{eqnarray}
\overline{x}^\mu =-\beta x^\mu /x^2  \label{eq4}
\end{eqnarray}

The inversion is singular on the light cone originating from the pole. To be
complete, one must consider two singular sets: the first one contains the
points in $\overline{{\cal R}}$ having their images in ${\cal R}$ at
infinity (characteristic equation $\overline{x}^2=0$) while the other one
contains the points in ${\cal R}$ which are images of points at infinity in $%
\overline{{\cal R}}$ (characteristic equation $x^2=0$). For the
transformations (\ref{eq3b}) to accelerated frames similarly, two singular
sets are defined by equations
\begin{eqnarray*}
1+2\alpha .\overline{x}+\alpha ^2\overline{x}^2 &=&0 \\
1-2\alpha .x+\alpha ^2x^2 &=&0
\end{eqnarray*}

In the present paper, we focus attention upon the problem of electromagnetic
fields in four-dimensional space-time. For the sake of comparison with
results obtained by Fulling and Davies \cite{FullingD} however, we
occasionally have a look onto the anomalous case of a two-dimensional
space-time. In this case, the conformal coordinate transformations defined
by equation (\ref{eq3b}) constitute a larger group than for other space-time
dimensions. The propagation equation of a massless scalar field is preserved
under arbitrary transformations of light-cone variables ($f_{-}$ and $f_{+}$
everywhere increasing functions)
\begin{eqnarray}
\overline{u}_{\pm } &=&\overline{x}^0\pm \overline{x}^1=f_{\pm }\left(
u\right)   \nonumber \\
u_{\pm } &=&x^0\pm x^1  \label{eq5}
\end{eqnarray}
But the transformations obeying also equation (\ref{eq2b}) are, in the
two-dimensional case as for other dimensions, the transformations generated
by Poincar\'{e} transformations and inversions and they correspond to
homographic functions
\begin{eqnarray}
&&f_{\pm }\left( u\right) =\frac{a_{\pm }u+b_{\pm }}{c_{\pm }u+d_{\pm }}
\nonumber \\
&&a_{\pm }d_{\pm }-b_{\pm }c_{\pm }=1  \label{eq6}
\end{eqnarray}

We may now discuss the invariance of uniformly accelerated motion in a
conformal coordinate transformation. The Abraham vectors $\overline{w}^\mu $
and $w^\mu $ defined according to definition (\ref{eq1}) for a given
trajectory in the inertial and conformal frames are related through
\begin{eqnarray*}
\overline{w}^\mu =\frac{\partial \overline{x}^\mu }{\partial x^\nu }\frac 1{%
\lambda ^3}\left\{ w^\nu +\left( v^\nu v^\rho -\eta ^{\nu \rho }\right)
v^\sigma \left( \varphi _{\rho \sigma }-\varphi _\rho \varphi _\sigma
\right) \right\}
\end{eqnarray*}
This coincides with equation (13) in ref. \cite{Hill} with minor misprints
corrected. As the conformal factor obeys equation (\ref{eq2b}), this
relation reduces to equation (15) in ref. \cite{Hill}
\begin{eqnarray*}
\overline{w}^\mu =\frac{\partial \overline{x}^\mu }{\partial x^\nu }\frac 1{%
\lambda ^3}w^\nu
\end{eqnarray*}

If the Abraham vector is null in a frame, it remains null after a conformal
coordinate transformation, so that uniformly accelerated motions are
preserved by conformal coordinate transformations. In particular, uniform
motion in the inertial frame $\overline{{\cal R}}$ (like rest for instance)
is transformed by the conformal coordinate transformations (\ref{eq3b}) into
a uniformly accelerated trajectory. This means that geodesic motion in the
conformal frame ${\cal R}$ is a uniformly accelerated motion. Conversely,
any uniformly accelerated motion in the inertial frame may be considered as
rest in a well-chosen conformal frame. In other words, conformal coordinate
transformations perfectly fit the definition of uniformly accelerated
motions for a point-like body. The extension to a body of finite size is
discussed in next section.

\section{Conformal transformations of light rays and distances}

Conformal transformations furthermore preserve uniform motion of massless
particles. This property can be considered a corollary of conformal
invariance of Maxwell equations \cite{Bateman}.

For making this point more explicit, we define a light ray by the equation
\begin{eqnarray*}
x^\mu  &=&x^{\prime \mu }+v^\mu \left( t-t^{\prime }\right)  \\
v^2 &=&0
\end{eqnarray*}
where $x_\mu ^{\prime }$ is a point on the ray, $v^\mu $ the light-like
4-velocity normalized so that $v^0=1$; $(t-t^{\prime })$ the coordinate time
along the ray. It is easily checked out \cite{Page} that this definition is
preserved by conformal coordinate transformations with the correspondances
\begin{eqnarray}
\overline{v}^\mu  &=&\frac{f_{\ \nu }^\mu \left( x^{\prime }\right) v^\nu }{%
f_{\ \nu }^0\left( x^{\prime }\right) v^\nu }  \nonumber \\
\overline{t}-\overline{t}^{\prime } &=&f_{\ \nu }^0\left( x^{\prime }\right)
v^\nu \lambda \left( x\right) \left( t-t^{\prime }\right)   \nonumber \\
\lambda \left( x\right)  &=&\frac \beta {1-2\alpha .x^{\prime }+\alpha
^2x^{\prime 2}-2\left( \alpha .v-\alpha ^2x^{\prime }.v\right) \left(
t-t^{\prime }\right) }  \nonumber \\
f_{\ \nu }^\mu \left( x\right)  &=&\frac 1{\lambda \left( x\right) }\frac{%
\partial \overline{x}^\mu }{\partial x^\nu }  \label{eq7}
\end{eqnarray}

Let us emphasize that the law of light propagation is preserved in conformal
coordinate transformations and that the singular sets cannot be considered
as horizons since they do not affect light propagation. In each frame
considered separately, the traversal of singular sets is not registered by
light. It appears however that the time intervals $t-t^{\prime }$ and $%
\overline{t}-\overline{t}^{\prime }$ measured along light rays in the two
reference frames are connected through an homographic relation which is
singular when the light ray crosses a singular set. The time intervals thus
undergo sign changes according to the law
\begin{eqnarray}
&&{\rm sgn}\left( t-t^{\prime }\right) ={\rm sgn}\left( \lambda \left(
x\right) \right) {\rm sgn}\left( \lambda \left( x^{\prime }\right) \right)
{\rm sgn}\left( \overline{t}-\overline{t}^{\prime }\right)   \nonumber \\
&&{\rm for\quad }\left( x-x^{\prime }\right) ^2=0  \label{eq8}
\end{eqnarray}
This law holds for the transformation (\ref{eq3b}) to accelerated frames and
for the inversion (\ref{eq4}) as soon as $\beta >0$; for $\beta <0$, global
sign has to be changed.

As a consequence of invariance of light rays, a Minkowski distance null in a
frame remains null in the other frame. For getting the transformation law of
non-null distances, we first notice that the inversion (\ref{eq4})
corresponds to a simple relation
\begin{eqnarray}
\left( \overline{x}-\overline{x}^{\prime }\right) ^2=\lambda \left( x\right)
\lambda \left( x^{\prime }\right) \left( x-x^{\prime }\right) ^2  \label{eq9}
\end{eqnarray}
where $\lambda $ is the conformal factor. As distance is preserved by
translations, this equation is still valid for the combined transformation (%
\ref{eq3b}) with the conformal factor (\ref{eq2b}).

In the Minkowskian space-time $\overline{{\cal R}}$,$(\overline{x}-\overline{%
x}^{\prime })^2$ represents a pseudo-Riemannian metric distance squared, and
equation (\ref{eq9}) appears as a relation between this metric invariant and
a squared distance $(x-x^{\prime })^2$ defined in the conformal frame ${\cal %
R}$ by using the Minkowskian metric rather than the conformal one. At the
limit of neighboring points, it reproduces the transformation law for the
metric tensor (from $\overline{g}_{\mu \nu }=\eta _{\mu \nu }$ in $\overline{%
{\cal R}}$ to $g_{\mu \nu }$ in ${\cal R}$ ). Equation (\ref{eq9}) may thus
be understood as an extension for finite distances of the characterization (%
\ref{eq2a}) of conformal transformations in terms of infinitesimal distances.

Usually, accelerated frames are represented by Rindler transformations which
fit uniformly accelerated motion of a point chosen as the origin while
preserving rigidity of spatially extended bodies \cite{Rindler}. The
conformal coordinate transformations used in the present paper do not
preserve rigidity, because of space-time variation of the scale factor, for
example in equations (\ref{eq2b}) and (\ref{eq9}). Rigidity is however
approximately preserved when the acceleration $a$ and the size $\delta $ of
the body obey the inequality
\begin{eqnarray*}
a\delta \ll c^2
\end{eqnarray*}
In the following, we consider that this condition is fulfilled. Uniformly
accelerated motion of a finite-size mirror may thus be identified with rest
in a well-chosen conformal frame.

\section{Definition of vacuum fluctuations}

Here, we recall the definition of vacuum fluctuations in Minkowski
space-time. Field fluctuations are characterized by the following
correlation functions (expectation values are evaluated for free fields in a
given state; $A$ and $B$ stand for field operators)
\begin{eqnarray*}
C_{AB}\left( x,x^{\prime }\right)  &\equiv &\left\langle A\left( x\right)
B\left( x^{\prime }\right) \right\rangle -\left\langle A\left( x\right)
\right\rangle \left\langle B\left( x^{\prime }\right) \right\rangle  \\
\sigma _{AB}\left( x,x^{\prime }\right)  &\equiv &\frac 1{2\hbar }\left\{
C_{AB}\left( x,x^{\prime }\right) +C_{BA}\left( x^{\prime },x\right)
\right\}  \\
\xi _{AB}\left( x,x^{\prime }\right)  &\equiv &\frac 1{2\hbar }\left\{
C_{AB}\left( x,x^{\prime }\right) -C_{BA}\left( x^{\prime },x\right)
\right\}
\end{eqnarray*}
Symmetrised $(\sigma _{AB})$ and antisymmetrised $(\xi _{AB})$ correlation
functions correspond respectively to field anticommutators and commutators.
The commutator correlation function is state independent and directly
related to the propagation equations; it describes the spectral density of
the field rather than its fluctuations. On the other hand, the
anticommutator correlation function characterizes fluctuations and depends
upon the field state.

Thermal states correspond to stationary correlation functions and may be
characterized by spectra defined as Fourier transforms of these functions
according to the general prescription
\begin{eqnarray*}
f\left( x,x^{\prime }\right) =\int \frac{d^4k}{\left( 2\pi \right) ^4}%
f\left[ k\right] e^{-ik\left( x-x^{\prime }\right) }
\end{eqnarray*}
Precisely, thermal equilibrium is defined through the
fluctuation-dissipation relations (we measure $T$ as an energy by including
the Boltzmann constant in its definition; $\omega $ is the frequency $k_0$)
\begin{eqnarray}
C_{AB}\left[ k\right]  &=&\frac{2\hbar \xi _{AB}\left[ k\right] }{%
1-e^{-\hbar \omega /T}}  \nonumber \\
\sigma _{AB}\left[ k\right]  &=&\coth \left( \frac{\hbar \omega }{2T}\right)
\xi _{AB}\left[ k\right]   \label{eq10}
\end{eqnarray}
This is just Planck relation between the spectral density of the field and
its fluctuations written in terms of quantum correlation functions \cite
{CallenW}. Vacuum is then defined as the limiting case of a zero
temperature, in which case only positive frequencies appear in the
correlation function $C_{AB}$
\begin{eqnarray}
C_{AB}\left[ k\right]  &=&2\hbar \theta \left( \omega \right) \xi
_{AB}\left[ k\right]   \nonumber \\
\sigma _{AB}\left[ k\right]  &=&{\rm sgn}\left( \omega \right) \xi
_{AB}\left[ k\right]   \label{eq11}
\end{eqnarray}
Here, $\theta \left( {}\right) $ is the Heaviside step function and ${\rm sgn%
}\left( {}\right) $ the sign function.

For a massless scalar field $\Phi $ in Minkowski space-time, these
correlation functions are easily determined from the propagation equation
\begin{eqnarray*}
\xi _{\Phi \Phi }\left[ k\right] =\pi {\rm sgn}\left( \omega \right) \delta
\left( k^2\right)
\end{eqnarray*}
Coming back from the momentum domain to the space-time domain, one obtains
the usual correlation function $C_{\Phi \Phi }$ proportional to ($t$ is the
time $x^0$; $\varepsilon \rightarrow 0^{+}$; we use natural space-time units
with $c=1$)
\begin{eqnarray}
c\left( x,x^{\prime }\right)  &=&\frac 1{\left( x-x^{\prime }\right)
^2-i\varepsilon \left( t-t^{\prime }\right) }  \nonumber \\
&=&{\cal P}\frac 1{\left( x-x^{\prime }\right) ^2} \nonumber \\
&&+i\pi \delta \left( \left(
x-x^{\prime }\right) ^2\right) {\rm sgn}\left( t-t^{\prime }\right)
\label{eq12}
\end{eqnarray}
The first term (${\cal P}$ stands for a principal part) corresponds to the
field anticommutator whereas the second term ($\delta $ represents the Dirac
function) corresponds to the commutator. The anticommutator differs from
zero for any couple of points whilst the commutator vanishes off the light
cone. In the electromagnetic case, the correlation functions of the
potential vector $A_\mu $ are found to be in Minkowski space-time
\begin{eqnarray}
C_{A_\mu A_\nu }\left( x,x^{\prime }\right) =\frac \hbar \pi \eta _{\mu \nu
}\ c\left( x,x^{\prime }\right)   \label{eq13}
\end{eqnarray}
We have used natural electromagnetic units with $4\pi \epsilon _0=1$ and
expressed potential vector in Feynmann gauge. The correlation functions of
the field tensor
\begin{eqnarray*}
F_{\mu \nu }=\partial _\mu A_\nu -\partial _\nu A_\mu
\end{eqnarray*}
are deduced from the former ones and are gauge independent. In the
following, we will consider that vacuum fluctuations refer to correlation
functions of such unambiguously defined physical quantities.

\section{Vacuum fluctuations in conformal accelerated frames}

As propagation equations are preserved by conformal transformations to
accelerated frames, the correlation functions describing vacuum fluctuations
are expected to be also preserved. We now prove that this is the case for
vacuum electromagnetic fluctuations.

We first consider the case of scalar fields, where vacuum fluctuations in
Minkowski frame $\overline{{\cal R}}$ are defined in the standard manner
described in the foregoing section; the correlation function $\overline{c}(%
\overline{x},\overline{x}^{\prime })$ is given by equation (\ref{eq12}) with
$x$ and $x^{\prime }$ replaced by $\overline{x}$ and $\overline{x}^{\prime }$%
. We deduce from equations (\ref{eq8}) and (\ref{eq9}) that this function is
transformed according to the simple law
\begin{eqnarray}
\overline{c}\left( \overline{x},\overline{x}^{\prime }\right) \lambda \left(
x\right) \lambda \left( x^{\prime }\right) &=&{\cal P}\frac 1{\left(
x-x^{\prime }\right) ^2} \nonumber \\
&&+i\pi \delta \left( \left( x-x^{\prime }\right)
^2\right) {\rm sgn}\left( t-t^{\prime }\right) \nonumber \\
&=&c(x,x^{\prime })
\label{eq14}
\end{eqnarray}
where $c\left( x,x^{\prime }\right) $ appears as a correlation function
defined in the conformal frame ${\cal R}$ in a Minkowskian manner.

Coming to the electromagnetic case, we start from the correlation functions
of the potential vector known in Minkowski frame $\overline{{\cal R}}$, i.e.
from equation (\ref{eq13}) with $x$ and $x^{\prime }$ replaced by $\overline{%
x}$ and $\overline{x}^{\prime }$. We transform the potential vector as a
covariant vector (using equation (\ref{eq7}))
\begin{eqnarray*}
A_\mu \left( x\right) =\frac{\partial \overline{x}^\rho }{\partial x^\mu }%
\left( x\right) \overline{A}_\rho \left( \overline{x}\right) =\lambda \left(
x\right) f_{\ \mu }^\rho \left( x\right) \overline{A}_\rho \left( \overline{x%
}\right)
\end{eqnarray*}
We then use equation (\ref{eq14}) to write the correlation function in the
conformal frame as
\begin{eqnarray*}
C_{A_\mu A_\nu }\left( x,x^{\prime }\right) =\frac \hbar \pi c\left(
x,x^{\prime }\right) f_{\ \mu }^\rho \left( x\right) \eta _{\rho \sigma \
}f_{\ \nu }^\sigma \left( x^{\prime }\right)
\end{eqnarray*}
Straightforward computations lead to
\begin{eqnarray*}
f_{\ \mu }^\rho \left( x\right) \eta _{\rho \sigma }f_{\ \nu }^\sigma \left(
x^{\prime }\right)  &=&\eta _{\mu \nu }+\varphi _\mu \left( x\right) \left(
x_\nu -x_v^{\prime }\right) \\
&&+\varphi _\nu \left( x^{\prime }\right) \left(
x_\mu ^{\prime }-x_\mu \right)  \\
&&-\frac 12\varphi _\mu \left( x\right) \varphi _\nu \left( x^{\prime
}\right) \left( x^{\prime }-x\right) ^2
\end{eqnarray*}
so that transformed vacuum fluctuations are characterized by
\begin{eqnarray*}
C_{A_\mu A_\nu }\left( x,x^{\prime }\right)  &=&\frac \hbar \pi \eta _{\mu
\nu }c\left( x,x^{\prime }\right)  \\
&&+\frac \hbar \pi \varphi _\mu \left( x\right) \left( x_\nu
-x_v^{\prime }\right) c\left( x,x^{\prime }\right)  \\
&&+\frac \hbar \pi \varphi _\nu \left( x^{\prime }\right) \left( x_\mu
^{\prime }-x_\mu \right) c\left( x,x^{\prime }\right)  \\
&&-\frac \hbar {2\pi }\varphi _\mu \left( x\right) \varphi _\nu \left(
x^{\prime }\right)
\end{eqnarray*}
The first line in right-hand side of this equation has exactly the same form
as Minkowskian correlation function (\ref{eq13}), while the second and third
lines appear as corrections. These corrections may however be eliminated by
a further gauge transformation of the potential vector, which is reminiscent
of the gauge transformation needed to obtain conformal invariance of Maxwell
equations written in terms of the potential vector \cite{Dirac}.

As a consequence, the corrections do not appear in the correlation functions
of the field tensor, resulting in the property that the correlation
functions for the transformed fields $F_{\mu \nu}$ have exactly the same
form in the conformal accelerated frame as the vacuum correlation functions
for the fields $\overline{F}_{\mu \nu}$ written in Minkowski frame.

At this point, we may emphasize that correlation functions are written in
the conformal accelerated frame in terms of a Minkowski metric, that is with
conventions differing from covariance conventions. The preservation of
vacuum fluctuations obtained in the present section must therefore be
distinguished from covariance and interpreted as a property of invariance
under conformal transformations, similar to invariance of Maxwell equations.

\section{Vacuum fluctuations after reflection upon a uniformly accelerated
mirror}

It is now easy to demonstrate that vacuum fluctuations are not modified by
reflection upon a uniformly accelerated mirror. As already discussed, we
define a uniformly accelerated trajectory in the inertial frame $\overline{%
{\cal R}}$ as the image through a coordinate transformation of rest in a
conformal frame ${\cal R}$.

We know that the correlation functions characterizing vacuum fluctuations
are unchanged in coordinate transformation from $\overline{{\cal R}}$ to $%
{\cal R}$. It is clear that they are also preserved by reflection upon the
motionless mirror in the conformal frame ${\cal R}$. They are still
preserved when coming back to the inertial frame $\overline{{\cal R}}$
through the inverse conformal coordinate transformation. We therefore deduce
that vacuum fluctuations are preserved by reflection upon a uniformly
accelerated mirror. More precisely, the correlation functions characterizing
the output fields have exactly the same form as those associated with the
input fields. Here, input and output fields are evaluated in the inertial
frame where there is no conformal factor. Note also that the invariance
property only holds for output correlation functions compared to input ones;
an analysis of cross correlations between input and output fields would
reveal the mirror's motion.

To summarize, reflection upon a uniformly accelerated mirror does not induce
any modification of vacuum and does not correspond to energy-momentum
radiation, as a consequence of invariance under conformal coordinate
transformations of vacuum fluctuations on one hand, of definition of
uniformly accelerated motion on the other hand.

Notice that, in the anomalous two-dimensional case, field commutators are
preserved, as the propagation equations, under all the coordinate
transformations of the large conformal group (\ref{eq5}). In contrast, field
anticommutators are preserved only under those transformations which obey
equation (\ref{eq9}), i.e. homographic conformal transformations (\ref{eq6}%
). The whole process of reflection upon a uniformly accelerated mirror is
also described by homographic conformal transformations and an exchange of
the two light-cone variables $u_\pm$, and invariance of vacuum under this
process obviously follows from this fact. For conformal transformations
which do not correspond to homographic functions, and hence do not fit
uniformly accelerated motions, vacuum fluctuations are modified, in
consistency \cite{QO92} with the emergence of a radiation reaction force for
a mirror moving with a non uniform acceleration \cite{FullingD}.

The relation between the absence of radiation from a uniformly accelerated
mirror and conformal invariance was already noticed for perfect mirrors
scattering vacuum fields in two-dimensional \cite{FullingD} or
four-dimensional \cite{FordV} space-time. It is indeed well-known that
reflection upon a perfect mirror constitutes a conformally-invariant
boundary condition \cite{CunninghamB}. One deduces from the result of the
present paper that the absence of radiation from a uniformly accelerated
scatterer holds more generally for scatterers which do not obey conformal
invariance. This is in particular the case for partially transmitting
mirrors where a frequency scale is given by the reflection cutoff \cite{QO92}%
, and for a Fabry-Perot cavity formed by two rigidly bound mirrors where a
length scale is given by their distance \cite{JP93b}.

\section{Discussion}

In the present paper, accelerated frames have been defined from inertial
frames through conformal coordinate transformations. We have verified that
field commutators are, as the propagation equations, preserved by such
transformations. We have shown that anticommutators are also preserved,
which means that the very definition of vacuum fluctuations as the zero
temperature state is the same in conformal accelerated frames as in inertial
frames. Using this invariance property, we have deduced that uniformly
accelerated mirrors in vacuum do not radiate energy-momentum.

This result raises questions about the precise physical significance of the
equivalence between accelerated vacuum and thermal state \cite{Unruh}.

Statements about vacuum fluctuations perceived in accelerated frames depend
upon the particular coordinate transformations chosen to define these
frames. In particular, vacuum fluctuations are preserved by conformal
coordinate transformations whilst they appear as thermal fluctuations after
Rindler transformations. Such a difference may not be surprising since
Maxwell equations are preserved by conformal transformations whilst they are
modified by Rindler hyperbolic transformations. In this respect, the
equivalence between acceleration and temperature has to be contrasted with
radiation from vacuum due to spacetime curvature created for example by
black holes \cite{Hawking}, which effect is clearlyunaffected by coordinate
transformations \cite{ZeldovichS}.

Now, statements about fluctuations reflected from mirrors with a uniformly
accelerated motion only depend upon the definition of such a motion, and not
upon a parametrization of accelerated frames used as an intermediate step in
the calculations. In the present paper, a conformal parametrization has been
used to establish the absence of radiation from a uniformly accelerated
mirror. A Rindler parametrization would have led to the same result.

As discussed in the introduction, symmetries of vacuum fluctuations are
directly connected to the problem of relativity of motion. Since vacuum
fluctuations are invariant not only under Lorentz transformations, but also
under conformal coordinate transformations to accelerated frames, we may
consider that this invariance explains why uniformly accelerated motions are
frictionless in vacuum, in the sense that they do not give rise to
radiation. When considering that the dissipative force experienced by the
mirror provides a mechanical detection of field fluctuations, it follows
that the same fluctuations are registered for uniformly accelerated or
inertial motions in vacuum. A more precise assessment of such a mechanical
detection technique could be obtained through the evaluation of the force
fluctuations experienced by uniformly accelerated or inertial mirrors.

\end{multicols}

\end{document}